\documentclass[apj,iop]{emulateapj}

\usepackage{natbib}
\usepackage{longtable}
\usepackage{hyperref}
\usepackage{amsmath}
\usepackage{graphicx}
\usepackage[dvipsnames]{xcolor}
\usepackage{booktabs}

\shorttitle{Luminous AGN in RGG 66}
\shortauthors{Kimbrell et al.}

\begin{document}

\title{A Luminous X-ray AGN in the Dwarf-Dwarf Galaxy Merger RGG 66}

\author{Seth J. Kimbrell}
\affil{Department of Physics and Astronomy, Carleton College, MN 55057}
\affil{eXtreme Gravity Institute, Department of Physics, Montana State University, MT 59715, USA}
\email{skimbrell@carleton.edu}

\author{Amy E. Reines}
\affil{eXtreme Gravity Institute, Department of Physics, Montana State University, MT 59715, USA}

\begin{abstract}
We present the discovery of a luminous X-ray AGN in the dwarf galaxy merger RGG 66.
The black hole is predicted to have a mass of $M_{\rm BH} \sim 10^{5.4} M_\odot$ and to be radiating close to its Eddington limit ($L_{\rm bol}/L_{\rm Edd} \sim 0.75$). The AGN in RGG 66 is notable both for its presence in a late-stage dwarf-dwarf merger and for its luminosity of 
$L_{\rm 2-10~keV} = 10^{42.2}$ erg s$^{-1}$, which is among the most powerful AGNs known in nearby dwarf galaxies. The X-ray spectrum has a best-fit photon index of $\Gamma = 2.4$ and an intrinsic absorption of $N_H \sim 10^{21}$ cm$^{-2}$. 
These results come from a follow-up {\it Chandra X-ray Observatory} study of four irregular/disturbed dwarf galaxies with evidence for hosting AGNs based on optical spectroscopy.  
 The remaining three dwarf galaxies do not have detectable X-ray sources with upper limits of $L_{\rm 2-10~ keV} \lesssim 10^{40}$ erg s$^{-1}$. 
 Taken at face value, our results on RGG 66 suggest that mergers may trigger the most luminous of AGNs in the dwarf galaxy regime, just as they are suspected to do in more massive galaxy mergers.

\end{abstract}

\section{Introduction}

In recent years, massive black holes (BHs) in dwarf galaxies ($M_{\rm BH} \sim 10^4-10^6~M_\odot$) have been increasingly discovered and studied in detail \citep[for a review, see][]{reines2022}. These systems provide clues to BH seeding as they give us an opportunity to study BHs which have not grown much compared to the supermassive BHs ($M_{\rm BH} \sim 10^6-10^9~M_\odot$) which are ubiquitous in more massive galaxies. At the same time, studying the demographics and morphologies of the dwarf galaxy hosts gives us a chance to characterize the environments inhabited by these BHs. Moreover, since current capabilities do not allow us to directly observe the first seed BHs in the early Universe, nearby dwarf galaxies hosting the smallest BHs offer our best chance put constraints on possible formation channels \citep{volonteri2021,greene2020,inayoshi2020,volonteri}.

Additionally, dwarf galaxy mergers are expected to be common in the earlier Universe and they have been shown to trigger periods of intense star-formation in the present-day Universe, leading to the formation of blue compact dwarf galaxies (BCDs) \citep{paudel2018,stierwalt2015}. However, active BHs in dwarf-dwarf mergers are an understudied area. The quintessential AGN in a dwarf-dwarf merger is Mrk 709 \citep{kimbro2021,reines2014}, a system in which an AGN has been detected in one of the members of an early-stage merger. \citet{micic2023} reported the discovery of strong candidates for the first dual AGNs in dwarf-dwarf mergers; systems in which the separation between the two AGNs are great enough to be resolvable (on the scale of kpc).

Here we present {\it Chandra X-ray Observatory} observations of four irregular/disturbed dwarf galaxies. Our sample comes from the work of \citet{kimbrell2021}, who used {\it Hubble Space Telescope} observations to study the structures and morphologies of dwarf galaxies that were selected by \citet{reines} as strong candidates for hosting active massive BHs based on optical spectroscopy from the Sloan Digital Sky Survey (SDSS). 

The primary focus of this work is the detection of a luminous X-ray AGN in the late-stage dwarf-dwarf merger, RGG 66 (ID 66 in the \citealt{reines} paper).  
In Section \ref{sec:sample} we describe our sample of dwarf galaxies in more detail. In Section \ref{sec:obs} we describe the X-ray observations. We present our analysis and results for RGG 66 in Section \ref{sec:analysis} and discuss the targets with non-detections in Section \ref{sec:nondetections}. We finish with our conclusions in Section \ref{sec:conclusion}.

\begin{figure*}[!h]
\includegraphics[width=\textwidth]{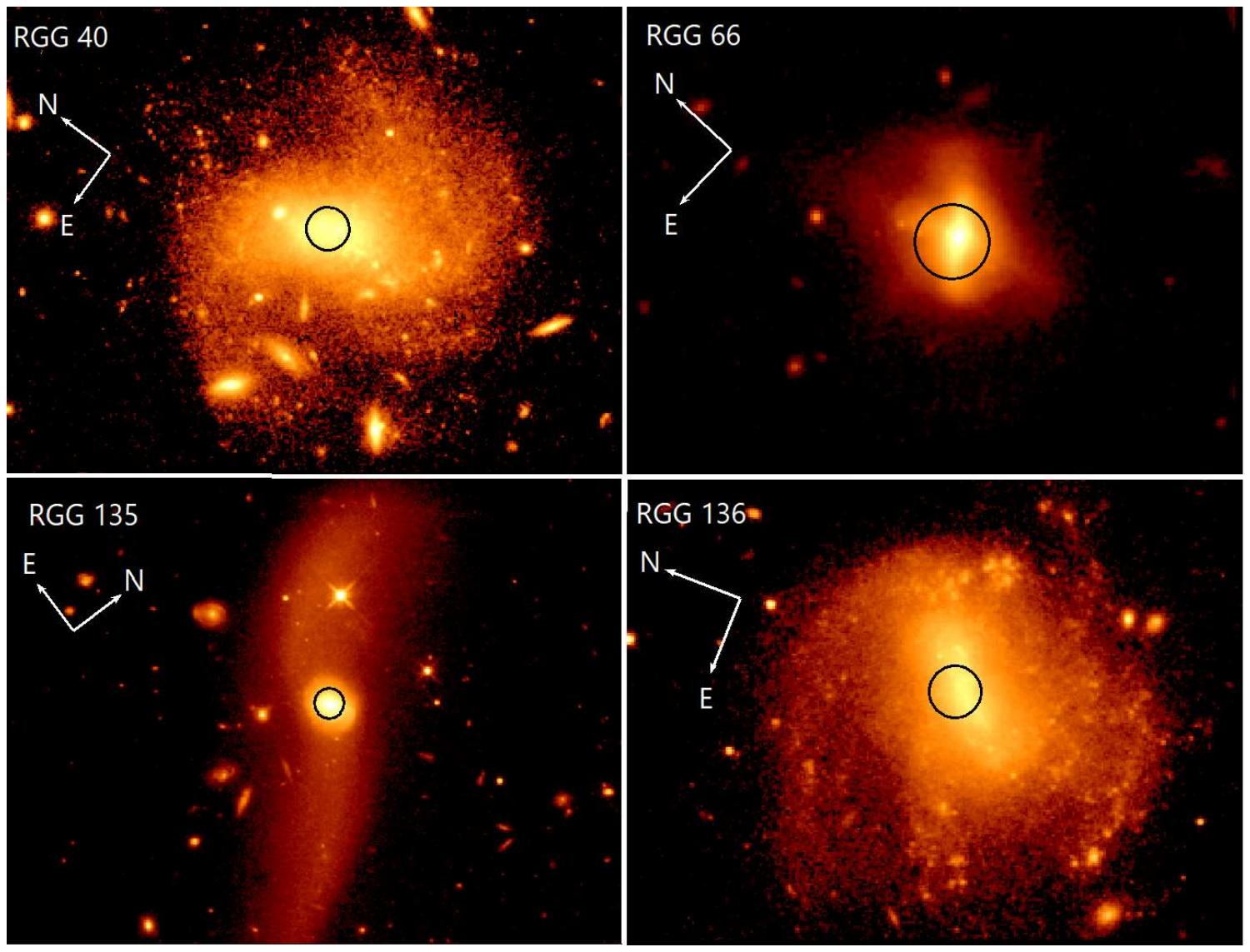}
\caption{ Near-IR observations from $HST$ \citep{kimbrell2021} of the four dwarf irregular galaxies analyzed in this work. The black circle indicates the 3\arcsec diameter SDSS spectroscopic fiber, with fiber positions obtained from the Nasa Sloan Atlas (NSA)
}
\label{fig:irregulars}
\end{figure*}

\begin{deluxetable*}{cccccccccc}[!htb]
\tabletypesize{\footnotesize}
\tablewidth{0pt}
\tablecaption{Irregular/Disturbed Dwarf Galaxy Sample \label{table:sample}}
\tablehead{
\colhead{RGG ID} & \colhead{NSAID} & \colhead{obsid} & \colhead{RA} & \colhead{Dec} & \colhead{$z$} & \colhead{r$_{\rm 50}$} & \colhead{log ($M_*/M_\odot$)} & \colhead{$N_{\rm H}$} & \colhead{Observation time} \\
\colhead {} & \colhead{} & \colhead{} & \colhead{(deg)} & \colhead{(deg)} & \colhead{} & \colhead{(kpc)} & \colhead{} & \colhead{($10^{20}$ cm$^{-2}$)} & \colhead{(ks)} \\ 
\colhead{(1)} & \colhead{(2)} & \colhead{(3)} & \colhead{(4)} & \colhead{(5)} & \colhead{(6)} & \colhead{(7)} & \colhead{(8)} & \colhead{(9)} & \colhead{(10)}} 
\startdata
RGG 40 & 82616 & 25280 & 117.12165 & 51.01453 & 0.0190 & 3.91 & 9.1 & 5.14 & 8.96 \\

RGG 66 & 55081 & 25281, 26315 & 154.44624 & 39.53551 & 0.0540 & 0.40 & 9.0 & 1.39 & 34.87$^a$ \\

RGG 135 & 4308 & 25282 & 263.01240 & 59.98194 & 0.0291 & 3.06 & 9.4 & 3.46 & 19.79 \\

RGG 136 & 5563 & 25283 & 359.03827 & $-$0.40800 & 0.0256 & 5.32 & 9.2 & 3.40 & 15.87

\enddata
\tablecomments{Column 1: ID given in \citet{reines} and \citet{kimbrell2021}. Column 2: NSA identification number. Column 3: Chandra Observation ID. Column 4: Right Ascension of the galaxy. Column 5: Declination of the galaxy. Column 6: Redshift taken from the NSA. Column 7: Petrosian 50\% light radius, from the NSA. Column 8: Log total stellar mass from the NSA. Column 9: Galactic neutral hydrogen column density. Column 10: Exposure time in kiloseconds.  \\ $^a$RGG 66 was observed twice - one observation of 18.79 ks and one of 16.08 ks, and we merged the observations for analysis.}
\end{deluxetable*}

\begin{figure*}[!t]
$\begin{array}{cc}
\hspace{.6cm}
{\includegraphics[width=0.47\textwidth]{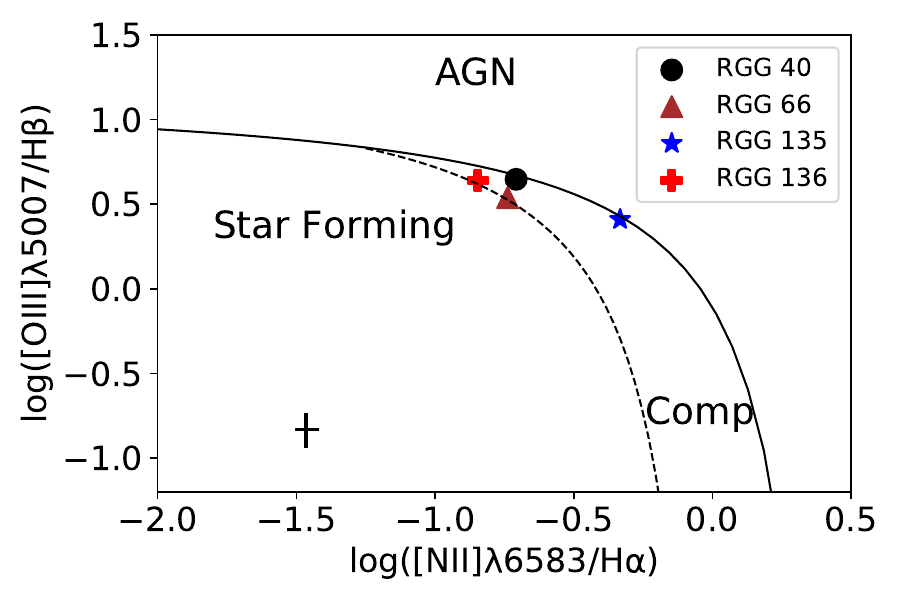}} &
{\includegraphics[width=0.47\textwidth]{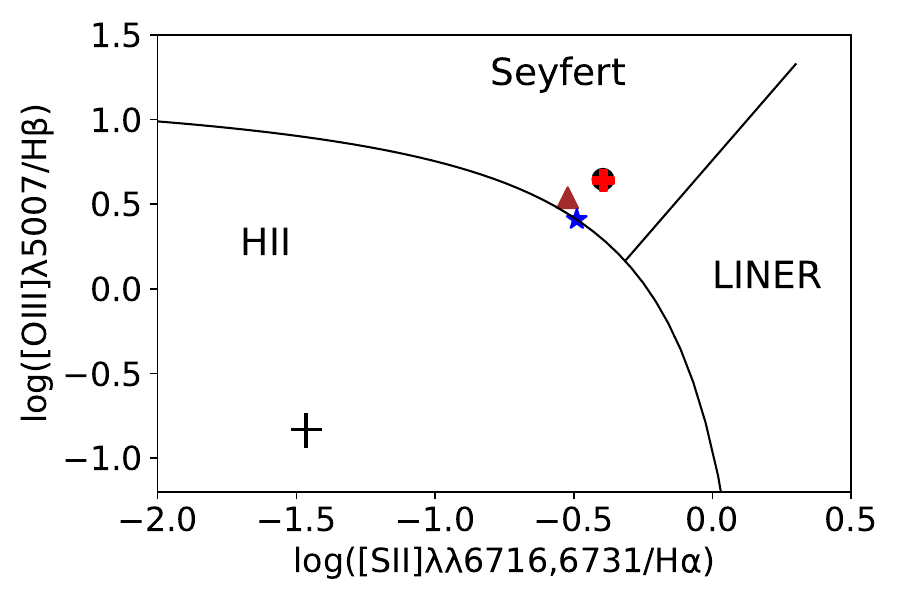}}
\end{array}$
\caption{Narrow emission line diagnostic diagrams for our sample of irregular dwarf galaxies, which were identified as displaying optical signatures of active massive BHs by \citet{reines}, and which were analyzed by \citet{kimbrell2021}. Left: [O${\rm III}$]/H$\beta$ vs.\ [N${\rm II}$]/H$\alpha$ diagnostic diagram, with the ``maximum starburst" line from stellar photoionization models \citep{KewleyStarburst} shown as the solid line, and the empirical separation from \citet{Kauffmann} between galaxies whose emission is dominated by star formation and galaxies with some contribution from AGN shown as the dashed line. Right: [O${\rm III}$]/H$\beta$ vs.\ [S${\rm II}$]/H$\alpha$ diagnostic diagram, adopting the classifications from \citet{kewleyactive}. Typical errors are shown in the lower left corners.}
\label{fig:BPT}
\end{figure*}

\section{Sample of Dwarf Galaxies}\label{sec:sample}
Our sample of disturbed/irregular dwarf galaxies comes from \citet{kimbrell2021}. In that work, \citet{kimbrell2021} analyzed {\it HST} near-IR imaging of a subsample of 41 dwarf galaxies which were identified by \citet{reines} as likely AGN hosts based on narrow emission line ratio diagnostics \citep{kewleyactive,Kauffmann,KewleyStarburst}. Six of the dwarf galaxies were classified as ``irregular/disturbed," broadly indicating that the galaxy could not be modeled by axisymmetric models. Of those six, three appeared to be Magellanic type dwarf irregulars, while two exhibited clear signatures of interactions or mergers. The final galaxy possessed internal spiral structure that made modeling difficult, but it did not fall into either the Magellanic-type irregular or disturbed category.

We excluded the galaxy with an internal spiral (RGG 53), as well as one of the Magellanic type irregulars (RGG 5), which had previously been observed by {\it Chandra} with no X-ray source detected. Of the four remaining galaxies (see Figure \ref{fig:irregulars}), two are Magellanic-type irregular dwarf galaxies (RGG 40 and RGG 136) and two show signs of interactions/mergers (RGG 66 and RGG 135). All four of these dwarf galaxies were identified as Seyferts in the [OIII/H$\beta$] vs.\ [SII]/H$\alpha$ diagnostic, and all fell into the Composite region of the [OIII]/H$\beta$ vs.\ [NII]/H$\alpha$ diagnostic (Figure \ref{fig:BPT}). We observed these four galaxies with {\it Chandra} and the galaxy properties are shown in Table \ref{table:sample}. Distances are obtained from the NASA Sloan Atlas assuming $h = 0.73$. Galactic neutral hydrogen column densities come from \citet{dickeylockman1990} and are retrieved from Chandra's Colden Galactic Neutral Hydrogen Density Calculator\footnote{https://cxc.harvard.edu/toolkit/colden.jsp}.

\section{Chandra X-Ray Observations}\label{sec:obs}

Our target galaxies were observed by {\it Chandra} between November 12, 2021 and January 19, 2023 with exposure times ranging from 9 ks to 35 ks (Table \ref{table:sample}). For each observation, the target galaxy was centered on the ACIS S3 chip.

Using version 4.14 of the Chandra Interactive Analysis of Observations software (CIAO) \citep{fruscione2006}, we first reprocessed our data utilizing the chandra\_repro script. 
We applied Chandra calibration files (CALDB 4.9.8) for reprocessing, filtered for any background flares, and created new event files which were used in our analysis.

We then attempted to correct the absolute astrometry of our images using the SDSS. We ran the CIAO wavdetect routine\footnote{https://cxc.cfa.harvard.edu/ciao/ahelp/wavdetect.html} on each filtered image using wavelets of size 1.0, 1.4, 2.0, 2.8 and 4.0 pixels. We set our significance threshold to be $10^{-6}$; this is the threshold at which we should expect roughly one strong background fluctuation to be detected as a source across the entire chip. If any X-ray sources were found outside the target galaxies, we matched them to existing SDSS detections. However, no matching sources were located for our images, and so astrometry corrections could not be performed in the end. 

\begin{figure}[h]
\begin{center}
\includegraphics[width=\linewidth]{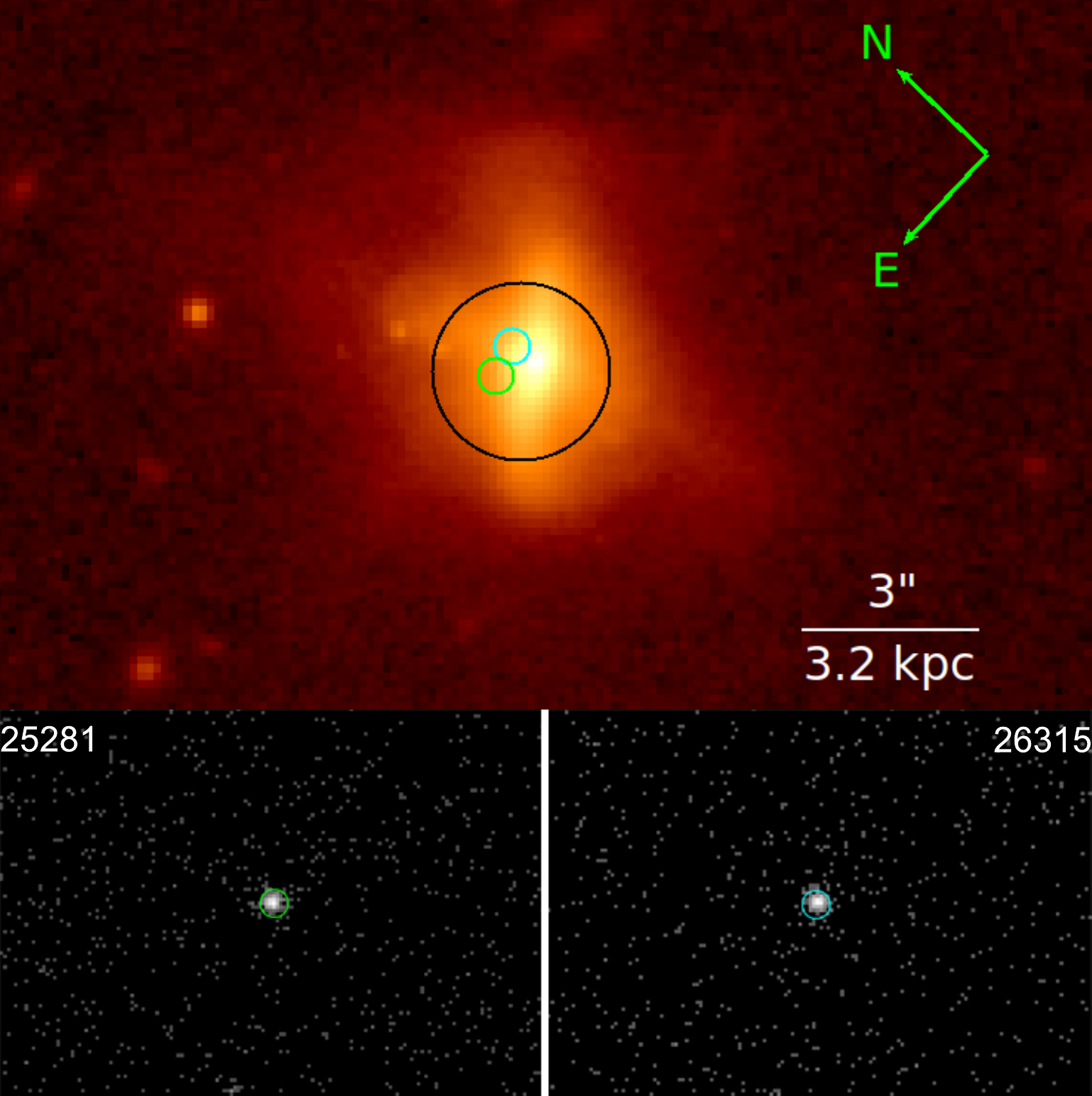}
\caption{Top: Near-infrared (F110W) {\it HST} image of RGG 66 shown on a log scale. The black circle shows the position of the SDSS spectroscopic fiber, while the smaller green and cyan circles show the position and positional uncertainty of the detected X-ray source for observations 25281 and 26315, respectively. The different positions of the X-ray source from the two observations is due to the pointing accuracy of {\it Chandra} rather than a real offset. Bottom: {\it Chandra} images for observations 25281 and 26315, with source apertures shown following the same color pattern as above.
}
\label{fig:rgg66}
\end{center}
\end{figure}

Next, we searched for X-ray point sources that could correspond to active massive BHs in our target galaxies. We began by filtering our images from 2 to 7 keV and running wavdetect. For our point-spread function map, we created a PSF map using the CIAO fluximage routine with an enclosed energy fraction of 39\% at 4 keV. Once wavdetect identified point sources, we filtered by location in the image and only accepted X-ray sources which are within 3r$_{50}$ of the center of the galaxy. We detect an X-ray source in one of our four target galaxies, RGG 66. We show the location of the X-ray source in RGG 66 along with the SDSS fiber position in the top panel of Figure \ref{fig:rgg66}.

For the other three galaxies, in which no sources were found by wavdetect, we determine upper limits on the fluxes/luminosities of potential X-ray sources at the 95\% confidence level via Poisson statistics using srcflux\footnote{https://cxc.cfa.harvard.edu/ciao/ahelp/srcflux.html}. We center a $4\arcsec$ radius circular source region at the aimpoint of the image, and we extract background counts using an annulus centered at the aimpoint of the image, with inner radius $4\arcsec$ and outer radius $20\arcsec$.  We assume an absorbed power-law spectral model using the Galactic $N_{\rm H}$ values towards each galaxy and a photon index of $\Gamma = 1.8$, a typical value for low-luminosity AGN \citep{latimer2021,ho2009,ho2008}. We find upper limits on the hard X-ray luminosities of $L_{2-10~{\rm keV}} \lesssim 10^{40}$ erg s$^{-1}$ for the three galaxies with non-detections (see Table \ref{table:sources}).

\begin{deluxetable*}{ccccccc}[]
\tabletypesize{\footnotesize}
\tablewidth{0pt}
\tablecaption{RGG 66 Observations \label{table:rgg66sources}}
\tablehead{
\colhead{ObsID}  & \colhead{Source RA} & \colhead{Source Dec} & \colhead{Source Pos. Error} & \colhead{Exp. Time} & \colhead{Net Counts (2-7 keV)} & \colhead{Count Rate (2-7 keV)}  \\
\colhead{} & \colhead{(deg)} & \colhead{(deg)} & \colhead{(arcsec)} & \colhead{(ks)} & \colhead{} & \colhead{(count/sec)}  \\
\colhead{(1)} & \colhead{(2)} & \colhead{(3)} & \colhead{(4)} & \colhead{(5)} & \colhead{(6)} & \colhead{(7)}}
\startdata
\cutinhead{Detected Source}
\\
25281 & 154.446325 & 39.535565 & 0.31 & 18.517 & 240.01 $\pm$ 27.43 & 0.0130  \\ \\

26315 & 154.446128 & 39.535606 & 0.31 & 16.066 & 209.52 $\pm$ 25.63 & 0.0130  \\
\enddata
\tablecomments{Column 1: Observation ID. Column 2: Right Ascension of the X-ray source. Column 3: Declination of the X-ray source. Column 4: 95\% error in position, not accounting for absolute astrometric uncertainties that are expected to be $\sim$1\arcsec. Column 5: Exposure time in kiloseconds. Column 6: Net Counts from 2-7 keV. Column 7: Count rate. 
}
\end{deluxetable*}

\begin{figure*}[t!]
\begin{center}
\includegraphics[width=5in]{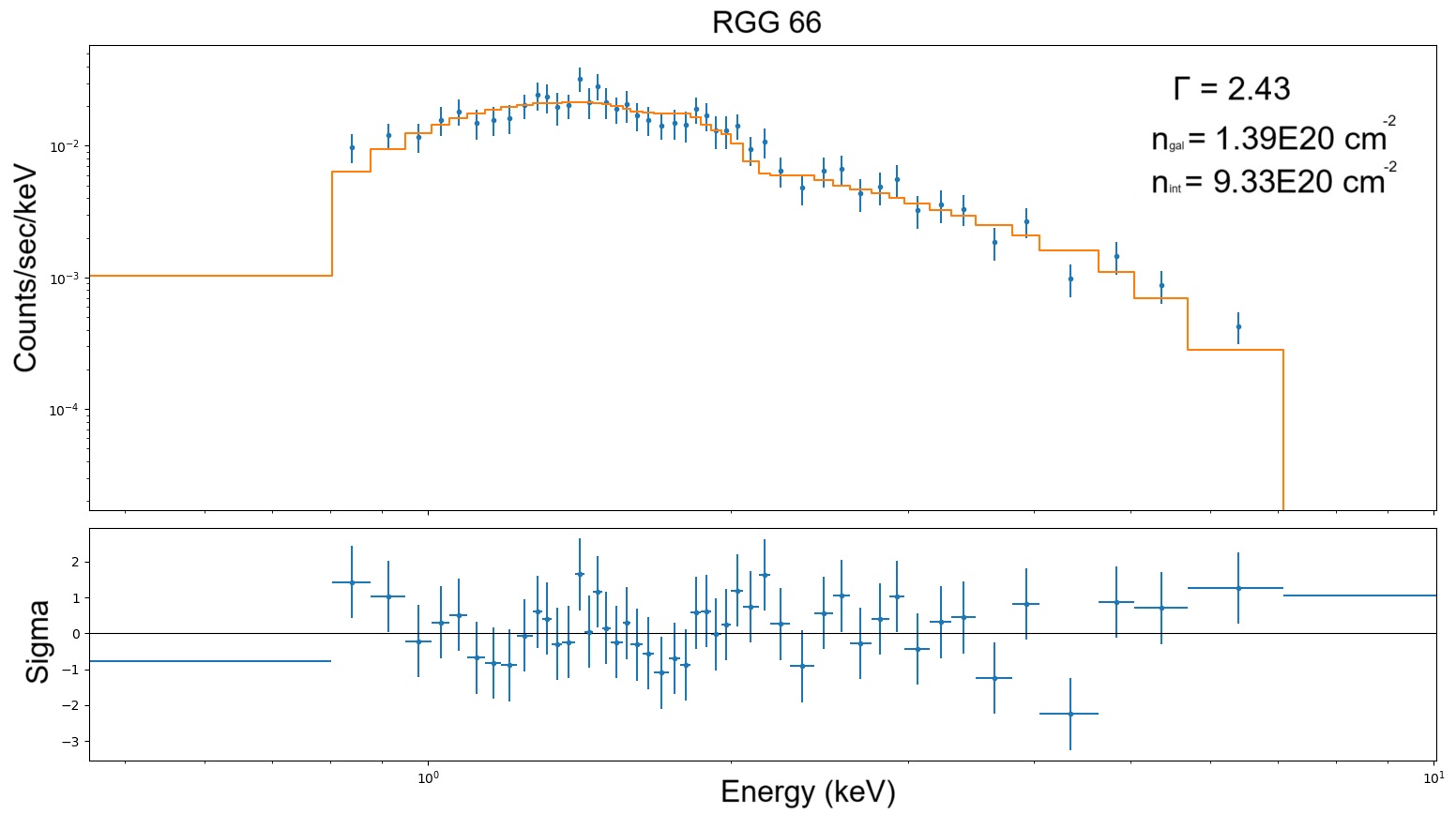}
\caption{ X-ray spectrum of RGG 66, with a best-fit photon index of $\Gamma = 2.43$ and counts grouped in bins of 20. 
}
\label{fig:rgg66_spec}
\end{center}
\end{figure*}

\section{A Luminous X-ray AGN in RGG 66}\label{sec:analysis}

\subsection{X-ray Source Properties}

We detected a bright X-ray source in each of the two observations of RGG 66 using wavdetect. The source is offset by $\sim 0\farcs65$ ($\sim 1.3$ pixels) in the two observations.  Given that the offset is within the absolute astrometric uncertainties of {\it Chandra} ($\sim$1\arcsec) and the count rates are the same (see below), we conclude that we are detecting the same source in the two images.  

The positions and positional uncertainties in each image are given in Table \ref{table:rgg66sources}.
 The positional uncertainty is given as the
 95\% error circle in the wavedetect source position using the relation from \citet{hong2005}, which depends on the offset from the aimpoint, $D$ (in arcminutes), and the net counts $c_n$:

\begin{equation}
    \begin{split}
    P_{err} = 0\farcs25 + \frac{0\farcs1}{{\rm log}(c_n + 1)}\left[1+\frac{1}{{\rm log}(c_n + 1)}\right] \\ + 0\farcs03\left[\frac{D}{{\rm log}(c_n + 2)}\right]^2 + 0\farcs0006\left[\frac{D}{{\rm log}(c_n + 3)}\right]^4.
    \end{split}
\end{equation}

\noindent
This does not account for any absolute astrometric uncertainty, which is expected to be on the order of $\sim$1\arcsec.

To find the net counts detected in each observation, we used a circular aperture centered on each detected source with a radius enclosing 90\% of the energy at 4.5 keV ($2\arcsec$). 
We then estimated the number of background counts by creating an annulus co-located with the source, with an inner radius equal to the radius of the source aperture and outer radius equal to 12 times the radius of the source aperture. We subtracted the background counts from the source counts to find the net counts and corrected for the 90\% enclosed energy fraction. 

We found net counts of $240.01 \pm 27.43$ ($\sim 0.013$ counts/sec) for the observation 25281 and net counts of $209.52 \pm 25.63$ ($\sim 0.013$ counts/sec) for observation 26315 in the 2-7 keV range. Since our source has net counts $\gg 10$, we neglected the background when calculating the errors in net counts and use the 90\% confidence intervals from \citet{gehrels1986}.

Given the large number of counts, we performed spectral analysis on the data for RGG 66. We used the specextract CIAO tool\footnote{https://cxc.cfa.harvard.edu/ciao/ahelp/specextract.html} to extract a spectrum from each observation using the same aperture and background used to find net counts, then merged the spectra using the combine\_spectra CIAO tool\footnote{https://cxc.cfa.harvard.edu/ciao/ahelp/combine\_spectra.html} (bottom panel of Figure \ref{fig:rgg66}). We grouped our counts in bins of 20 then used the Sherpa fitting package to model the spectrum with an absorbed power-law model, including a Galactic absorption term  $N_{H, gal}$ and an intrinsic absorption term  $N_{H, target}$. We froze $N_{H,gal}$ to the value of $1.39 \times 10^{20}~{\rm cm^{-2}}$ from \citet{dickeylockman1990}. We found that the spectrum was best fit by an absorbed power law with photon index $\Gamma = 2.43 \pm 0.14$ and intrinsic absorption of $N_{H, target} = (9.3 \pm 6.7) \times 10^{20}~{\rm cm^{-2}}$ (see Figure \ref{fig:rgg66_spec}). We find an unabsorbed flux in the 2-10 keV range of $2.59 \times 10^{-13}$ erg/s. 

The hard X-ray luminosity of the source in RGG 66 is $L_{\rm 2-10~keV} = 10^{42.18}$ erg s$^{-1}$, squarely in the AGN regime and more than $\sim 1000\times$ the expected contribution from stellar-mass X-ray binaries (see \S\ref{sec:host} and Table \ref{table:sources}).

\subsection{BH Mass and Eddington Ratio}

We first determine the minimum mass of the BH in RGG 66 by assuming Eddington-limited accretion. While the Eddington ratio may in fact be lower than 1, this calculation will place a lower limit on the BH mass. 
The Eddington luminosity is given by

\begin{equation}
    L_{\rm Edd} \sim 1.26 \times 10^{38} \left( \frac{M_{{\rm BH}}}{M_\odot} \right).
\end{equation}

\noindent
and to determine the minimum BH mass, we assume $L_{\rm Edd} = L_{\rm bol} = \kappa L_{\rm 2-10 keV}$. We use the relation derived in \citet{duras2020} to find the bolometric correction, $\kappa$, from the hard X-ray luminosity:

\begin{equation}
    \kappa(L_x) = a\left[1 + \left( \frac{log(L_x/L_\odot)}{b} \right) ^c \right]
\end{equation}
\noindent with best-fit values of $a = 15.33 \pm 0.06$, $b = 11.48 \pm 0.01$, and $c = 16.20 \pm 0.16$. Using this relation and our measured hard X-ray luminosity, we estimate a bolometric correction of $\kappa \sim 15.47$. We then have $L_{\rm Edd} = L_{\rm bol} = \kappa L_{\rm 2-10 keV} = 
2.34 \times 10^{43}$ erg s$^{-1}$, corresponding to a minimum BH mass of $M_{\rm BH} \sim 10^{5.3}~M_\odot$.

Next, we estimate the BH mass using the relation in \citet{reinesvolonteri2015} between BH mass and total galaxy stellar mass for local AGNs:

\begin{equation}
    {\rm log}(M_{\rm BH}/M_\odot) = \alpha + \beta {\rm log}(M_{\rm stellar}/10^{11}M_\odot)
\end{equation}
\noindent with $\alpha = 7.45 \pm 0.08$ and $\beta = 1.05 \pm 0.11$. 
Applying this relation to RGG 66 with a total stellar mass estimate of $M_{\rm stellar} \sim 10^9~M_\odot$ (see Table \ref{table:sample}), we predict a BH of mass $10^{5.4}M_\odot$ with an uncertainty of $\sim 0.55$ dex \citep{reinesvolonteri2015}.

We can find the corresponding Eddington ratio of the BH in RGG 66 using this mass estimate and the hard X-ray luminosity of $L_{\rm 2-10~keV} = 10^{42.18}$ erg s$^{-1}$, 
where the Eddington ratio is given by

\begin{equation}
    f_{\rm Edd} = \frac{\kappa \times (L_{\rm 2-10 keV})}{L_{\rm Edd}}.    
\end{equation}

\noindent
This gives an Eddington ratio of $f_{\rm Edd} = 0.75^{+1.88}_{-0.54}$. The uncertainties come from propagating the uncertainty of 0.55 dex in the BH mass estimate using the relation from \citet{reinesvolonteri2015}, which we expect to dominate the error in Eddington ratio. Our results provide evidence that RGG 66 hosts a BH with a mass of a few $\times 10^5~M_\odot$ radiating at a high fraction of its Eddington luminosity.

\subsection{Host Galaxy}
\label{sec:host}

We first examine the properties of RGG 66 using the NASA-Sloan Atlas (NSA) to confirm its classification as a dwarf galaxy system. To begin, its Petrosian 50\% light radius of $r_{50} = 0.4$ kpc speaks to a compact system. We also compare the absolute magnitudes of RGG 66 and compare them to those found in literature for known dwarf galaxies. The AGN-hosting dwarf-dwarf merger system Mrk 709 \citep{reines2014,kimbro2021} has absolute $g, r$ and $i$ band magnitudes of $\sim -20$. For comparison, RGG 66 has absolute magnitudes of $\sim -18.2, -18.7$ and $-18.5$ in the $g, r$ and $i$ band. The $g$ band magnitude of RGG 66 is comparable with the rest of the \citet{reines} sample of dwarf galaxies from which it was drawn. These parameters, along with its stellar mass estimate of $M_\star \sim 10^9~M_\odot$, allow us to confidently describe RGG 66 as a dwarf.

We use visual inspection of RGG 66 in classifying it as a merger. The $HST$ image (top panel of Figure \ref{fig:rgg66}) shows tidal features running left to right in the image. These tidal tails are indicators of a galaxy merger/interaction. The presence of these features without a companion galaxy in the vicinity leads us to classify RGG 66 as a late-stage merger near the point of coalescence.

We estimate the star formation rate of RGG 66 using far-UV and mid-IR luminosity measurements from the {\it Galaxy Evolution Explorer (GALEX)} All-Sky Catalog and the {\it Wide-field Infrared Explorer (WISE)}:

\begin{equation}
    L(FUV)_{corr} = L(FUV)_{\rm obs} + 3.89L(25\mu m)
\end{equation}

\begin{equation}
    {\rm log}\dot M({\rm M_\odot yr^{-1})} = {\rm log}L(FUV_{\rm corr}) - 43.35
\end{equation}
\citep{kennicutt2012,hao2011}. The GALEX FUV luminosity is obtained via measurements in the NASA Sloan Atlas. While the calibrations above utilized 25$\mu m$ luminosities from the Infrared Astronomical Satellite ({\it IRAS}), we use 22$\mu m$ observations from {\it WISE} as the flux density ratio at these wavelengths is of order unity \citep{jarrett2013}. For RGG 66, with log L(FUV)[erg/s] = 42.48 and log L(22$\mu m$)[erg/s] = 42.08, 
we expect an SFR of 0.34 M$_\odot$ yr$^{-1}$ (assuming the mid-IR and FUV luminosities are dominated by star formation and any contribution from the AGN is negligible).

The corresponding specific star formation rate is 3.45 $\times 10^{-10}$ yr$^{-1}$.
At this specific star formation rate, we expect to see a contribution to the galaxy-wide X-ray luminosity from high-mass X-ray binaries (HMXB), which scale with star formation rate, with additional contributions from low-mass X-ray binaries (LMXB), which scale with galaxy mass \citep{mineo2012,lehmer2010,colbert2004,grimm2003}. 
 
We use the following relation from \citet{lehmer2010} to estimate the contribution from both HMXBs and LMXBs:
\begin{equation}
    L_{{\rm HX}}^{{\rm gal}} = \alpha M_* + \beta {\rm SFR}
\end{equation}
\noindent where $\alpha = (9.05 \pm 0.37) \times 10^{28}$ ergs s$^{-1} M_\odot^{-1}$ and $\beta = (1.62 \pm 0.22) \times 10^{39}$ ergs s$^{-1} (M_\odot {\rm yr}^{-1})^{-1}$ (with a $1\sigma$ scatter of 0.34 dex). For RGG 66, the expected 2-10 keV luminosity from X-ray binaries is $\sim 10^{38.8}$ erg s$^{-1}$. This is more than three orders of magnitude below the measured hard X-ray luminosity, indicating that the X-ray source in RGG 66 is due to an AGN.

The AGN in RGG 66 is also detected at radio wavelengths by the Very Large Array Sky Survey (VLASS). Eberhard et al.\ (submitted) present a search for radio AGNs in dwarf galaxies using VLASS and identify this object in their sample (ID 4 in that work). The radio source has a luminosity of $L_{\rm 3 GHz} = 10^{21.9}$ W Hz$^{-1}$.

\subsection{Comparison to Other AGNs in Dwarf Galaxies}\label{sec:comparison}
 
\begin{figure*}[!t]
\begin{center}
{\includegraphics[width=0.8\textwidth]{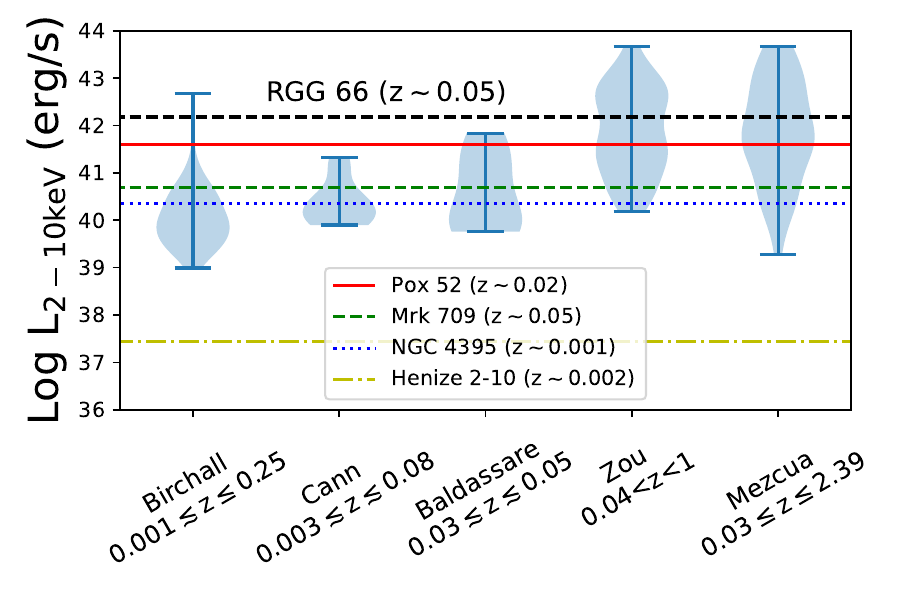}} 
\caption{X-ray luminosities in the 2-10 keV band of the source in RGG 66 compared to the sources studied in  \citet{cann2024,zou2023,birchall2020,mezcua2018,baldassare2017xray}, as well as Pox 52 and NGC 4395 \citep{dewangan2008}, along with Mrk 709 \citep{reines2014} and Henize 2-10 \citep{reines2016}. The range of luminosities for each sample is shown using the blue lines, while the width of the blue shaded region denotes the fraction of the sample at that luminosity. Luminosities have been transformed to the 2-10 keV range where necessary (see \S \ref{sec:comparison})}
\label{fig:comparelums}
\end{center}
\end{figure*}

We compare the X-ray luminosity of the active BH in RGG 66 to other low-mass AGNs in dwarf galaxies. Using the spectrum of RGG 66, we find a broad-band X-ray luminosity of log(L$_{0.5-8 {\rm keV}}) = 42.18$, significantly higher than the majority of known AGNs in nearby dwarf galaxies with X-ray observations (Figure \ref{fig:comparelums}).

Two of the most well-studied dwarf galaxies hosting optically-selected AGNs are NGC 4395 \citep{filippenko} and Pox 52 \citep{Barth}. While NGC 4395 is variable in the X-ray \citep{moran2005,lira1999},  measurements of its hard X-ray luminosity have generally been found to be  on the order of L$_{\rm 2-10 keV} \sim 10^{40} {\rm erg/s}$ \citep{dewangan2008,vaughan2005,moran2005}. Pox 52 was found by \citet{dewangan2008} to have a 2-10 keV luminosity of $\sim 10^{41.61}$ using XMM-Newton observations. 
The nearby (d $\sim$ 9 Mpc) dwarf starburst galaxy Henize 2-10 also hosts a massive BH at its center. This was observed using {\it Chandra} by \citet{reines2016} and found to have a 0.3-10 keV luminosity $\sim 10^{38}$ erg/s. Another massive black hole resides in the dwarf galaxy pair Mrk 709 \citep{reines2014}, which was observed by {\it Chandra} in that work and found to have a 2-10 keV luminosity of $10^{40.7}$ erg/s.

A number of studies have also focused on the X-ray properties of larger samples of AGNs in dwarf galaxies. For example, the 10 broad-line AGNs in the \citet{reines} sample of dwarf galaxies with optical AGN signatures have 2-10 keV X-ray luminosities in the range $L_{\rm 2-10 keV} = 10^{39.8}$ to $10^{41.8}$ erg/s \citep{baldassare2017xray}.

\citet{latimer2019} performed a combined X-ray and radio search among blue compact dwarf galaxies (BCDs) and located one candidate AGN, in Haro 9. That AGN candidate had a 2-10 keV luminosity of ${\rm L_{2-10 keV}} = 10^{39.4} {\rm erg/s}$. In a later work, \citet{latimer2021} performed an X-ray study of WISE-selected AGN candidates in dwarf galaxies. The five galaxies which they identified as having strong evidence for an accreting central BH had hard X-ray luminosities in the range $10^{40.1} - 10^{41.9} {\rm erg/s}$, with a median of $10^{40.3} {\rm erg/s}$.

\citet{birchall2020} presented a study of 61 X-ray selected AGNs in nearby (z $\leq$ 0.25) dwarf galaxies with X-ray luminosities given in the 2-12 keV band. RGG 66 has an X-ray luminosity of 10$^{42.18}$ erg/s, which is at the very high end of the range measured there. \citet{birchall2020} found 2-12 keV X-ray luminosities between $10^{39.05}$ erg/s and $10^{42.73}$ erg/s, with a median of $10^{40.04}$ erg/s (see Figure 4 of that work). In fact, RGG 66 has a higher hard X-ray luminosity than all but one dwarf galaxy reported in Table B1 of that work. 

\citet{zou2023} searched for active dwarf galaxies in the XMM-Spitzer Extragalactic Representative Volume Survey (XMM-SERVS). In that work, 73 active dwarf galaxies with redshift z $\leq$ 1 had their X-ray properties measured; X-ray luminosities in the 2-10 keV band ranged from $10^{40.18}$ to $10^{43.66} {\rm erg/s}$, with a median of $10^{41.85} {\rm erg/s}$ and a mean of $10^{41.86} {\rm erg/s}$.

\citet{cann2024} explored archival XMM observations of 37 low-metallicity dwarf galaxies and examined their X-ray properties. In that work, 10 galaxies were defined as candidate AGN hosts using the criterion that the X-ray luminosity in the 0.3-10 keV range exceeded $10^{40}$ erg/s. These 10 had full-band X-ray luminosities ranging from $10^{40.13}$ to $10^{41.56}$ erg/s.

\citet{mezcua2018} searched the {\it Chandra} COSMOS-Legacy survey and found 40 dwarf galaxies hosting AGNs at redshifts as far out as $z \sim 2.4$. In that work, X-ray luminosities were reported in the 0.5 - 10 keV range and fell in the range of $10^{39.5}$ to $10^{43.9}$ erg/s, with a median of $10^{42}$ erg/s and a mean of $10^{42.08}$ erg/s. Similar to RGG 66 in this work, many of the AGNs studied in \citet{mezcua2018} are accreting at very high Eddington ratios.

The massive BH in RGG 66 shares similarities with those found in \citet{micic2023}. In that work, two dwarf-dwarf merger systems with redshifts $\sim 0.27$ were identified as candidate dual AGNs (i.e., AGNs resolvable as two separate sources). In particular, the AGNs with pair separations of $< 5$ kpc in the galaxies named Elstir and Vinteuil have broad-band luminosities of log(L$_{0.5-8 {\rm keV}}) = 41.96$ and $42.71$ in erg s$^{-1}$, which are similar to that of the AGN in RGG 66. Unlike RGG 66, which appears to be a system at the coalescence stage of the merger, Elstir and Vinteuil appear to be in the very early stages of merging.

In a follow-up work, \citet{micic2024} used {\it HST} and {\it Chandra} imaging to identify six more AGNs in dwarf-dwarf mergers. The AGNs presented in that work have luminosities in the 0.3 - 7 keV band ranging from $10^{40.06}$ to $10^{43.56}$ erg/s. Five of the six systems have log(L) $< 10^{42}$ erg/s, and the remaining source is an outlier with log(L) $= 10^{43.56}$ erg/s.

Figure \ref{fig:comparelums} shows the 2-10 keV luminosities of the aforementioned BHs along with RGG 66. Luminosities which were not reported in the 2-10 keV energy band were converted from their reported energy band using the Portable, Interactive Multi-Mission Simulator (PIMMS)\footnote{https://cxc.harvard.edu/toolkit/pimms.jsp}. We assumed a photon index of $\Gamma = 1.8$ and N$_{\rm H}$ = N$_{\rm H,GAL}$. Some of the above BHs had spectral analysis performed or had fluxes obtained from a catalog which assumed values of $\Gamma$ and/or N$_{\rm H}$; in this case, we used the values used in the work.

The X-ray source in RGG 66 has a higher luminosity than any other comparably local source discussed above. While RGG 66 lies at z = 0.054, the next closest object with a 2-10 keV luminosity $\geq 10^{42.18}$ is in the \citet{zou2023} sample and lies at a redshift of 0.22.

\begin{deluxetable*}{ccccccccc}[htb]
\tabletypesize{\footnotesize}
\tablewidth{0pt}
\tablecaption{X-Ray Sources \label{table:sources}}
\tablehead{
\colhead{Name}  & \colhead{RA} & \colhead{Dec} & \multicolumn{2}{c}{\underline{Net Counts}} & \multicolumn{2}{c}{\underline{Flux ($10^{-15}$ erg s$^{-1}$ cm$^{-2}$)}} & \multicolumn{2}{c}{\underline{Luminosity (log(erg s$^{-1}$))}} \\
\colhead{} & \colhead{(deg)} & \colhead{(deg)} & \colhead{0.5-2 keV} & \colhead{2-7 keV} & \colhead{0.5-2 keV} & \colhead{2-10 keV} & \colhead{0.5-2 keV} & \colhead{2-10 keV} \\
\colhead{(1)} & \colhead{(2)} & \colhead{(3)} & \colhead{(4)} & \colhead{(5)} & \colhead{(6)} & \colhead{(7)} & \colhead{(8)} & \colhead{(9)}}
\startdata
\cutinhead{Detected Source}
\\
RGG 66 & 154.446334 & 39.535563 & 853.73 $\pm$ 51.24 & 465.01 $ \pm$ 36.00 & 436.08 & 259.08 & 42.42 & 42.18 \\
\cutinhead{Upper Limits on Non-Detections}
\\
RGG 40 & - & - & - & - & $<4.71$ & $<11.0$ & $<39.54 $& $<39.90$\\ \\

RGG 135 & - & - & - & - & $<4.84$ & $<12.3$ & $<39.92$ & $<40.32$\\ \\

RGG 136 & - & - & - & - & $<3.59$ & $<8.56$ & $<39.68$ & $<40.05$ \\

\enddata
\tablecomments{Column 1: Galaxy ID. Column 2: Right Ascension of the X-ray source, identified by wavdetect. Column 3: Declination of the X-ray source, identified by wavdetect. Columns 4-5: Aperture-corrected net counts from the combined spectrum. Columns 6-7: Absorption-corrected flux. Columns 8-9: Absorption-corrected log luminosity.}
\end{deluxetable*}

\section{Targets with Non-Detections}
\label{sec:nondetections}

The three galaxies without X-ray detections have upper limits on their 2-10 keV luminosities of $L_{2-10~{\rm keV}} \lesssim 10^{40}$ erg s$^{-1}$ (\S\ref{sec:obs}). Here we compare these upper limits to expectations based on multiwavelength AGN scaling relations for more massive galaxies. In particular, we compare our results to the relationships between $L_X$, L$_{{\rm H}\alpha}$ and L$_{\rm OIII}$ from \citet{panessa2006}:

\begin{equation}
    {\rm log} (L_X) = (1.06 \pm 0.04){\rm log}(L_{{\rm H}\alpha}) + (-1.14 \pm 1.78)
\end{equation}

\begin{equation}
    {\rm log} (L_X) = (1.22 \pm 0.06){\rm log}(L_{[{\rm OIII}]}) + (-7.34 \pm 2.53)
\end{equation}

\noindent where the luminosities are in erg/s. These relations come from a sample of 47 Seyfert galaxies in the Palomar optical spectroscopic survey of nearby galaxies \citep{ho1997}. \citet{latimer2021} estimated the scatter in the \citet{panessa2006} relations and found $\sim 0.72$ dex for the L$_{\rm X}$ - L$_{{\rm H} \alpha}$ relation, and a scatter of $\sim 0.66$ dex for the L$_{\rm X}$ - L$_{\rm OIII}$ relation.

\begin{figure*}[!t]
\begin{center}
$\begin{array}{cc}
{\includegraphics[width=0.5\textwidth]{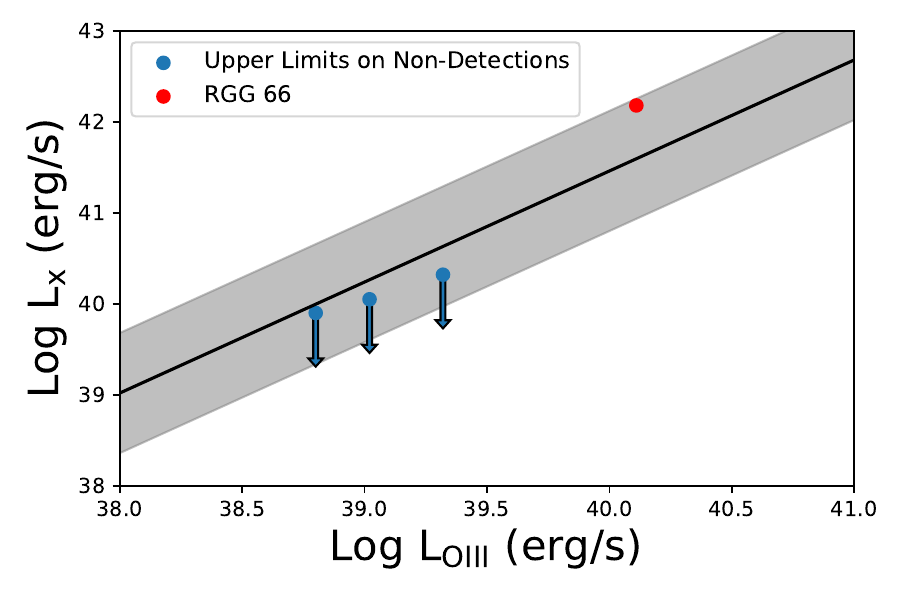}} &
{\includegraphics[width=0.5\textwidth]{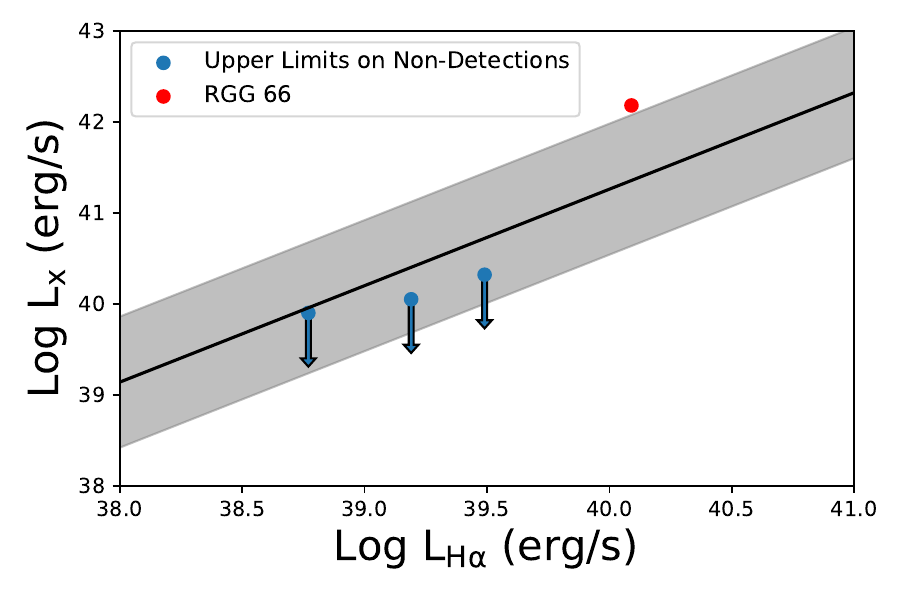}} 
\end{array}$
\caption{Observed and upper limit 2-10 keV luminosity vs observed O${\rm III}$ luminosity (top) and H$_\alpha$ luminosity (bottom) in erg/s. The black line shows the \citet{panessa2006} relation and the gray shading shows the scatter derived in \citet{latimer2021}.}
\label{fig:xraypredicted}
\end{center}
\end{figure*}

We obtain [OIII] and H$\alpha$ luminosities from \citet{reines} and use these to predict 2-10 keV luminosities. Figure \ref{fig:xraypredicted} shows that the upper limits fall within the scatter of both relations and the non-detections do not rule out the presence of low-luminosity AGNs given their optical line strengths. Moreover, there is a growing body of evidence that AGNs in dwarf galaxies tend to be less bright in X-rays than expected from typical scaling relations derived from more massive galaxies hosting AGNs, with possible explanations for the low X-ray luminosities including obscuration and intrinsic X-ray weakness \citep[e.g.,][]{latimer2021,arcodia2024}. In any case, it is quite possible that RGG 40, RGG 135 and RGG 136, which all have optical line ratios supporting the case for AGNs, host AGNs with low enough X-ray luminosities to be undetected by this search.


\section{Conclusions}\label{sec:conclusion}

We have presented {\it Chandra} observations of four low-mass irregular dwarf galaxies which were optically selected (via narrow-line diagnostic diagrams) as hosts of active massive BHs \citep{kimbrell2021,reines}. We have detected a luminous AGN in the late-stage dwarf-dwarf merger RGG 66. 
This is one of the first AGNs detected in such a late-stage merger of dwarf galaxies. A summary of our results is given below. 

\begin{enumerate}
  \item The X-ray source detected in RGG 66 is almost certainly an AGN, as the observed hard X-ray luminosity of $L_{\rm 2-10 keV} = 10^{42.18}$ erg s$^{-1}$ is nearly three orders of magnitude higher than that expected from X-ray binaries given the host galaxy star formation rate and stellar mass.
  \item The corresponding minimum BH mass estimated assuming Eddington-limit accretion is $M_{\rm BH} = 10^{5.3}M_\odot$.  
  Using the BH mass - total stellar mass relation of \citet{reinesvolonteri2015}, we predict a BH of mass $M_{\rm BH} \sim 10^{5.4}M_\odot$. The corresponding Eddington ratio is $f_{Edd} \sim 0.75$. These results indicate that the BH has a mass of a few hundred thousand solar masses and is radiating close to its Eddington limit.
  \item The X-ray spectrum of the AGN in RGG 66 is best fit by an absorbed power-law model with a photon index of  $\Gamma = 2.43 \pm 0.14$ and intrinsic absorption of $N_{H, target} = (9.3 \pm 6.7) \times 10^{20}~{\rm cm^{-2}}$.
  \item The remaining three irregular/disturbed dwarf galaxies with optically-selected AGNs in our sample are not detected in X-rays with upper limits of $L_{\rm 2-10~ keV} \lesssim 10^{40}$ erg s$^{-1}$. While X-ray detections would have helped confirm the presence of accreting massive BHs in these galaxies, the lack of detectable X-ray emission does not rule out the presence of massive BHs.
\end{enumerate}

The active BH in RGG 66 is also notable for being on the upper ends of the distributions of luminosity and Eddington ratio among known AGNs in dwarf galaxies. In fact, it is one of the brightest known AGNs in a dwarf galaxy at such low redshift.  This may be connected to its presence in a galaxy merger, which can be conducive to efficiently fueling central massive BHs. 

Indeed, simulations suggest that at least for higher mass systems, the brightest AGNs are located in galactic mergers \citep{hopkins2009,hopkins2008}. Observations suggest similar results. \citet{comerford2015} performed an observational study of 12 dual-AGN candidates and suggested that mergers tend to host more luminous AGNs. A similar result was found by \citet{treister2012}, who performed a multi-wavelength study of AGNs across a wide range of luminosities and redshifts, and also found mergers to host the most luminous AGNs. While a small sample size, our results on RGG 66 taken at face value, along with results from \citet{micic2023} suggest that the trend of mergers hosting the most luminous AGNs may extend to the low-mass regime as well.

\acknowledgements

The authors thank Lily Latimer for being a constant source of assistance with X-ray analysis. Support for this work was provided by NASA through Chandra Award Number GO2-23078X issued by the Chandra X-ray Observatory Center, which is operated by the Smithsonian Astrophysical Observatory for and on behalf of the NASA under contract NAS8-03060. AER also acknowledges support provided by NASA through EPSCoR grant number 80NSSC20M0231 and the NSF through CAREER award 2235277. This work has made use of software provided by the Chandra X-ray Center (CXC) in the application packages CIAO and Sherpa. This paper employs a list of Chandra datasets, obtained by the Chandra X-ray Observatory, contained in~\dataset[DOI: 10.25574/cdc.238]{https://doi.org/10.25574/cdc.238}.

\bibliography{refs}

\end{document}